\documentclass[preprint]{aastex}
\usepackage{color}

\shorttitle{Orbit of $\beta$ Pic b}
\shortauthors{Nielsen et al.}

\begin{document}

\title{The Gemini NICI Planet-Finding Campaign: \\ The Orbit of the Young Exoplanet $\beta$ Pictoris b}

\author{Eric L. Nielsen,\altaffilmark{1}
Michael C. Liu,\altaffilmark{1}
Zahed Wahhaj,\altaffilmark{2}
Beth A. Biller,\altaffilmark{3}
Thomas L. Hayward,\altaffilmark{4}
Jared R. Males,\altaffilmark{5}
Laird M. Close,\altaffilmark{5}
Katie M. Morzinski, \altaffilmark{5}
Andrew J. Skemer,\altaffilmark{5}
\\
Marc J. Kuchner,\altaffilmark{6}
Timothy J. Rodigas, \altaffilmark{5}
Philip M. Hinz, \altaffilmark{5}
Mark Chun,\altaffilmark{1}
Christ Ftaclas,\altaffilmark{1}
Douglas W. Toomey\altaffilmark{7}
}

\altaffiltext{1}{Institute for Astronomy, University of Hawaii, 2680
Woodlawn Drive, Honolulu, HI 96822, USA}
\altaffiltext{2}{European Southern Observatory, Alonso de Cordova 3107, 
Vitacura, Casilla 19001, Santiago, Chile}
\altaffiltext{3}{Institute for Astronomy, The University of Edinburgh, 
Royal Observatory, Blackford Hill, Edinburgh EH9 3HJ, UK}
\altaffiltext{4}{Gemini Observatory, Southern Operations Center, c/o AURA,
Casilla 603, La Serena, Chile}
\altaffiltext{5}{Steward Observatory, University of Arizona, 933 North Cherry
Avenue, Tucson, AZ 85721, USA}
\altaffiltext{6}{Goddard Space Flight Center, Code 667, Greenbelt, MD 20771, USA}
\altaffiltext{7}{Mauna Kea Infrared, LLC, 21 Pookela St., Hilo, HI
96720, USA}

\begin{abstract}
We present new astrometry for the young (12--21 Myr) 
exoplanet $\beta$~Pictoris~b taken 
with the Gemini/NICI and Magellan/MagAO 
instruments between 2009 and 2012.  
The high dynamic range of our observations allows us to measure the 
relative position of $\beta$ Pic b with respect to its primary star with 
greater accuracy than previous observations.  
Based on a Markov Chain Monte Carlo analysis, we find the planet has 
an orbital semi-major axis of 
9.1$^{+5.3}_{-0.5}$ AU and orbital eccentricity 
$<0.15$ at 68\% confidence (with 95\% confidence intervals of 8.2--48 AU and 
0.00--0.82 for semi-major axis and eccentricity, respectively, due to a 
long narrow degenerate tail between the two).  
We find that the planet has reached its maximum projected 
elongation, enabling higher precision determination of the orbital 
parameters than previously possible, and that the planet's projected 
separation is currently decreasing.  With unsaturated data of the 
entire $\beta$ Pic system (primary star, planet, and disk) obtained 
thanks to NICI's semi-transparent focal plane mask, we are able to 
tightly constrain the relative orientation of the 
circumstellar components.  We find the orbital 
plane of the planet lies between the inner and outer disks: 
the position angle (PA) of nodes for the planet's orbit 
(211.8$\pm$0.3$^\circ$) is 7.4$\sigma$ 
greater than the PA of the spine of 
the outer disk and 3.2$\sigma$ less than the warped inner disk PA, indicating 
the disk is not collisionally relaxed.  Finally, 
for the first time we are able to dynamically constrain the mass of 
the primary star $\beta$~Pic to 1.76$^{+0.18}_{-0.17}$ M$_{\sun}$.
\end{abstract}


\section{Introduction}

$\beta$ Pic is a young ($\sim$12--21 Myr, 
\citealt{barrado99,zsbw01,binks14}), 
nearby (19.44 pc, \citealt{newhip}) A6 star that hosts one of the most 
prominent known debris disks 
(e.g. \citealt{betapicdisk,wahhaj03,weinberger03,golimowski06,lagrange12}).  
The disk midplane is warped, with a 4$^\circ$ offset between the 
inner warped disk and the outer main disk, suggesting a 
giant planet influencing the disk \citep{mouillet97}.  This planet 
$\beta$~Pic~b was first detected in data from 2003, and the planet 
reappeared on the other side of the star in 2009 \citep{betapicb,betapic2}.  
$\beta$ Pic b is one of the first planets to be directly imaged and has the 
smallest projected physical separation of any imaged planet to date.  With 
a contrast of 9 magnitudes at $K_S$-band and a 
current projected separation of $\sim$0.4'', the planet is challenging to detect even with state-of-the-art 
adaptive optics.

Orbital properties of directly imaged planets can encode clues to 
their formation. The eccentricities of these planets, for example, may trace 
migration or planet-planet interactions 
(e.g., \citealt{takeda05,juric08,wang11}).  $\beta$ Pic b 
represents the 
longest-period exoplanet whose full orbital parameters can be determined with 
present observations.  
The estimated orbital period of $\sim$20 years allows us the 
opportunity to determine the orbit of this planet, whereas most other directly 
imaged planets will require many more 
decades of observations for a robust orbit determination.  
The orbital parameters of $\beta$ Pic b 
are of particular interest since they allow us to study the relationship 
between the planet and the debris disk.

The Gemini NICI Planet-Finding Campaign was a 4-year survey to detect 
extrasolar planets conducted between 2008 and 2012 
\citep{liunici,niciastars,debris,moving_groups}.  In addition to 
detecting a number of brown dwarf companions \citep{pztel,cd35,hd1160}, we 
detected the planet $\beta$ Pic b at multiple epochs over the course of the 
Campaign.  We combine these observations with new data from 
Magellan MagAO and previous work to determine the orbit of the 
planet.

\section{Observations}

\textit{VLT/NACO:} \citet{chauvin_orbit} present 9 epochs of 
$\beta$ Pic b astrometry from VLT NACO, between 2003 and 2011.  
\citet{bonnefoy_orbit} present an additional epoch from January 2012, as 
well as an improved orbital fit.  We use the 
10 astrometric observations and errors as reported in these works.

\textit{Gemini-South/NICI:} We observed $\beta$ Pic b with NICI at 
Gemini-South at 4 epochs between 2009 and 2012 (Table~\ref{table1}).  
Reductions were performed with 
the Gemini NICI Planet-Finding Campaign pipeline, described in depth in
\citet{pipeline}.  These data are discussed in detail in 
\citet{jared_betapic}, which also includes a comparison to an independent 
analysis of some of the data by \citet{boccaletti13}.  
We find that for overlapping epochs there is good 
agreement between the NICI and NACO astrometry, indicating that there is 
no significant astrometric offset between the two instruments.  In addition, 
NICI measurements have smaller uncertainties compared to those from 
NACO: the median separation and PA uncertainties are 0.005'' and 
0.6$^\circ$ for NICI and 0.011'' and 1.81$^\circ$ for NACO.  This 
increased precision is due to the partially transparent 
focal plane mask of NICI, which allows for an accurate measurement of the 
location of the star relative to the planet.  
\citet{bonnefoy_orbit} note the primary source of error 
in their astrometry is the uncertainty in the position of the star.

\textit{Magellan/MagAO}: 
$\beta$ Pic b was observed on UT 2012 December 04 in the $Y_S$ 
(0.985 $\mu$m) filter of Magellan MagAO+VisAO, and on UT 
2012 December 01, 02, 04, and 
07 in the 3.1 $\mu$m, 3.3 $\mu$m, $L'$, and $M'$ filters of Magellan 
MagAO+Clio2.  The astrometry and photometry from MagAO+VisAO is described 
in detail in \citet{jared_betapic} and from MagAO+Clio2 in 
\citet{katie_betapic}.  Similar to NICI, the Magellan MagAO+VisAO instrument 
has a partially transmissive focal plane mask so the primary star is 
unsaturated in the data.  MagAO+Clio2 data in 3.1 $\mu$m, 3.3 $\mu$m, and 
$L'$ were taken with long and short exposures of saturated and unsaturated 
data, while $M'$ data were entirely short unsaturated exposures.

The VisAO astrometry calibration is tied to 
the wider field of view Clio2 camera, and the two cameras are mounted 
simultaneously on the same rotator so they will have some common systematics, 
but the observations of the planet by these two cameras are otherwise 
independent measurements.  To calibrate the Clio2 
astrometry, we observed the Trapezium Cluster in the Orion Nebula as a 
reference.  Over the nights of 2012 Dec. 3rd, 4th, 6th, and 8th UT, we 
locked the AO loop on either $\theta ^1$ Ori B or C.  Stars $\theta ^1$ 
Ori A and E were also in the fields, as were many fainter Trapezium stars 
(see \citealt{katie_betapic} for details).  We compared the positions of the 
stars to their positions in \citet{close12} to derive the 
platescale, instrument angle, and distortion solution.  We applied these 
solutions to our $\beta$ Pic data to calibrate them.  The position of the 
planet was then measured in the Clio2 images by a grid search over (x,y)
positions.  

\section{MCMC Orbit Fitting}

\subsection{Methods}

We use a Metropolis-Hastings Markov Chain Monte Carlo approach to fit orbital 
parameters to the astrometric motion of $\beta$ Pic b from 2003 to 
2012, following the procedure of \citet{ford05} and \citet{ford06}.  
Such an approach has previously been applied to 
astrometric orbits by, e.g., \citet{2mass1534} and \citet{dupuy10} for 
substellar companions, as well as 
by \citet{chauvin_orbit} and \citet{bonnefoy_orbit} for $\beta$ Pic b 
itself.  
We compute two types of fits: one with the total system 
mass fixed (as done by all previous published orbital fits) and another 
with the total mass as a 
free parameter.  For the fixed-mass case, we have six free parameters: 
the semi-major axis ($a$), 
eccentricity ($e$), inclination angle ($i$), argument of periastron 
($\omega$), position angle of nodes ($\Omega$), and the 
epoch of periastron passage ($T_0$).  The period ($P$) is determined from the 
semi-major axis using Kepler's Third Law and adopting a total system mass of 
1.75 M$_{\sun}$ \citep{crifo_betapic}.  
For the floating-mass fit, the orbital period is the seventh 
free parameter, so the total mass is allowed to take on any value.

At the start of the chain, initial values are chosen for each free 
parameter.  A proposed trial step is taken by choosing a displacement in all 
parameters 
by randomly drawing from six (or seven) Gaussians centered on the current 
parameters with fixed standard deviations.  The probability 
ratio between current and proposed orbit parameters 
is computed by determining the difference in the $\chi^2$ statistic 
($\Delta \chi^2$) for all 
astrometric epochs for $\beta$~Pic~b between the initial 
and trial sets of parameters.  The choice of whether to adopt the 
trial parameters as the new step or to retain the current set is made via the 
Metropolis Hastings algorithm with probability ratio 
$\propto e^\frac{- \Delta \chi^2 }{ 2}$.  
Non-physical parameters ($a\le$ 0, $P\le$ 0, 
$e<$ 0, $e\ge$ 1) are assigned a $\chi^2$ of 10$^6$ while the three viewing 
angles ($i$, $\omega$, $\Omega$) and epoch of periastron passage ($T_0$) 
are allowed to take any values without 
limits.  To ensure the chains run efficiently, the standard 
deviations of the Gaussians for 
choosing trial steps are chosen initially via an iterative procedure 
where chains are run with different values of standard deviation 
for each parameter.  The standard deviations 
corresponding to an acceptance rate of 0.25 are used for the 
final chains \citep{dupuy10}.  

After we terminate the MCMC chain, the angles are then wrapped around 
between 0$^\circ$ and 180$^\circ$ for inclination angle, 
-180$^\circ$ and 180$^\circ$ for argument of periastron, and 
0$^\circ$ and 360$^\circ$ for position angle of nodes.  Epoch of 
periastron passage is wrapped around to give a value in the range 
[2005,2005+period].  2005 is chosen to provide a 
single distribution of epoch of periastron passage free of 
discontinuities.  Without radial velocity measurements of the star's 
reflex motion 
there is an ambiguity between $\Omega$ and $\Omega + 180^\circ$ --- that is, 
whether the south/eastern half of the orbit is closer to the Earth or 
the north/western half --- so while our chains only explored parameter space 
near $\Omega$=212$^\circ$, the results are equally valid for 
$\Omega$=32$^\circ$.\footnote{\citet{lagrange_rv} present HARPS radial 
velocity data of $\beta$ Pic between 2003 and 2010.  The 
RV induced by the activity of the star dominates the expected reflex motion: 
the measured RV varies between $-1.0$ and 0.66 km/s with 
a standard deviation of 0.28 km/s, while the expected RV variation between 
2003 and 2011 for the reflex motion of a 
10 M$_{Jup}$ planet with the median orbital parameters from the 
MCMC fit is 0.13~km/s.  The 
linear fit to the RV computed by \citet{lagrange_rv} indicates an 
increasing RV between 2003 and 2010, consistent with $\Omega$ near 212.  
This orientation is confirmed by the recent measurement 
of the RV of the planet by \citet{snellen14}, finding it to be moving 
toward the Earth at -15.4$\pm$1.7 km/s on 17 December 2013 UT.}

Ten chains are run from the same starting 
parameters so that we may test for convergence using the Gelman-Rubin (GR) 
statistic, with 10$^8$ steps per chain and parameters saved every 1000 
steps.  The GR statistic is 
essentially the ratio of the variance in each 
parameter in individual chains to the total variance for all chains.  
If each chain is sampling the same region of parameter space then the GR 
metric will be very close to unity.  If the chains are still exploring 
different regions of parameter space when the chains are terminated, then the 
GR statistic will be significantly larger than 1.  
A GR statistic 
less than 1.1 indicates the chains are converging, while less than 1.01 means 
excellent convergence \citep{ford06}.  

\subsection{Results}

We present medians and confidence intervals for each parameter in the 
fixed-mass case in Table~\ref{table2}.  The inclination 
angle, argument of periastron, position angle of nodes, and epoch of 
periastron passage have a GR 
statistic less than 1.01, and the other two parameters have a GR statistic 
less than 1.1, indicating that our 
posterior distributions for all parameters are reliable.

\begin{figure}
\epsscale{0.9}
\plotone{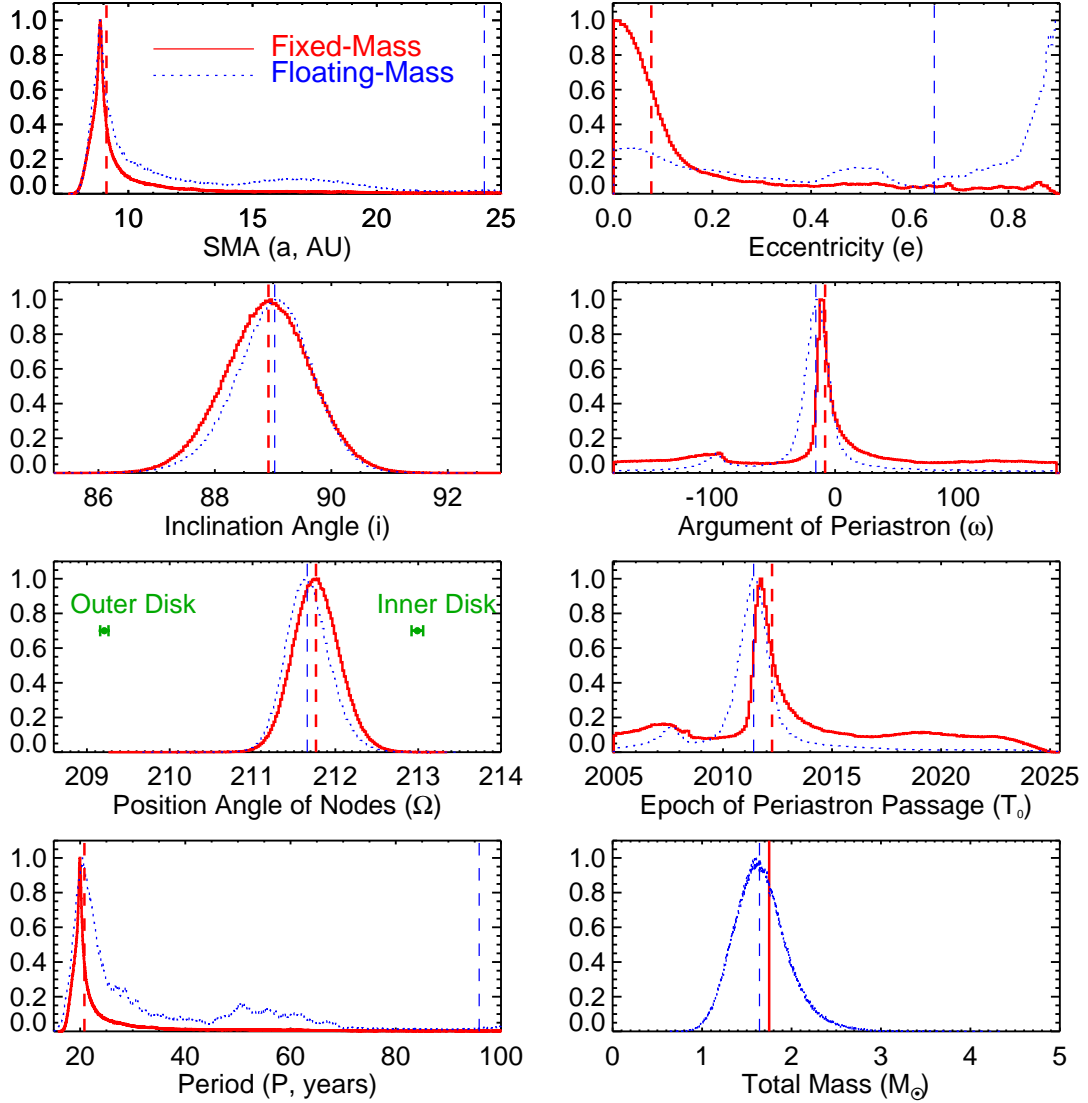}
\caption{Marginalized distributions of orbital parameters from our MCMC 
fit to the orbit of $\beta$ Pic b, for the cases where the total mass of the 
system is fixed at 1.75 M$_{\sun}$ (red solid lines) 
and floating as a free parameter (blue dotted lines).  The distributions have 
been normalized so that their peaks are unity.  Dashed lines 
mark the median of each distribution, and green points and error bars 
denote the position angle of the outer main disk and inner warped disk, 
without the overall astrometric calibration error of 
0.2$^\circ$ that is common to 
the measurement of the planet and the two disks.  Since the orbital fit draws 
on data from different instruments, the PDF for position angle of nodes 
does include the absolute astrometric uncertainties, though (as expected) the 
uncertainty on the PDF is less than the uncertainty for the PA measurement 
at any epoch.
\label{histfig}}
\end{figure}

\begin{figure}
\epsscale{0.9}
\plotone{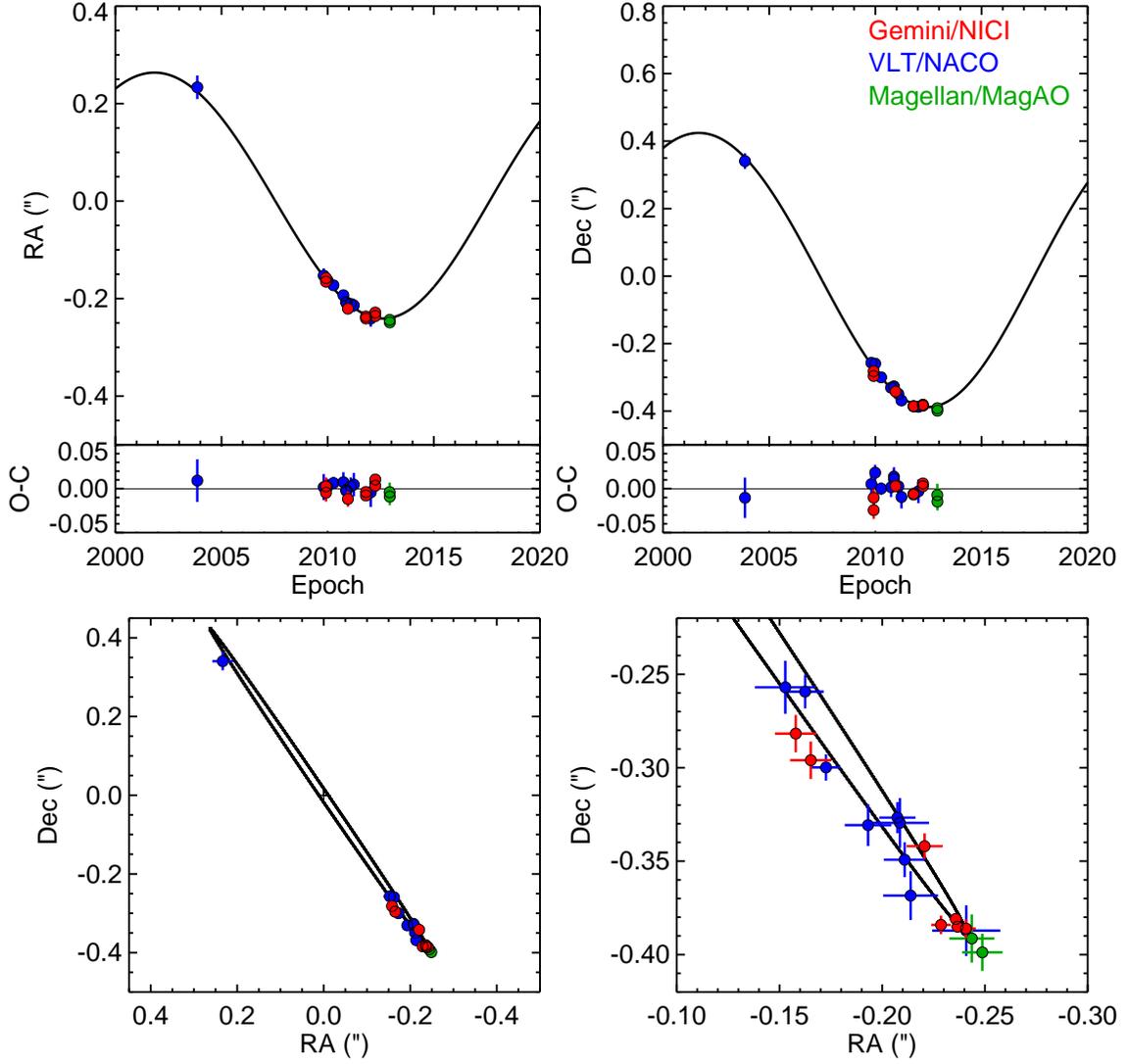}
\caption{The lowest $\chi^2$ orbit within the 68\% confidence interval 
from our fixed-mass MCMC chain, and 
residuals plotted as Observed - Calculated (O-C), 
corresponding to a $\chi^2_\nu$=1.38 for 31 degrees of freedom.  
\label{orbitfig}}
\end{figure}

Figure~\ref{histfig} displays the resulting marginalized posterior probability 
distributions for the orbital parameters and total system mass for 
both the fixed-mass and floating-mass cases.  Figure~\ref{orbitfig} shows 
an example set of orbital parameters from the fixed-mass MCMC chains that 
has the lowest $\chi^2$ 
within the 68\% confidence region for all parameters 
($\chi_\nu^2$=1.38).

\begin{figure}
\epsscale{0.9}
\plotone{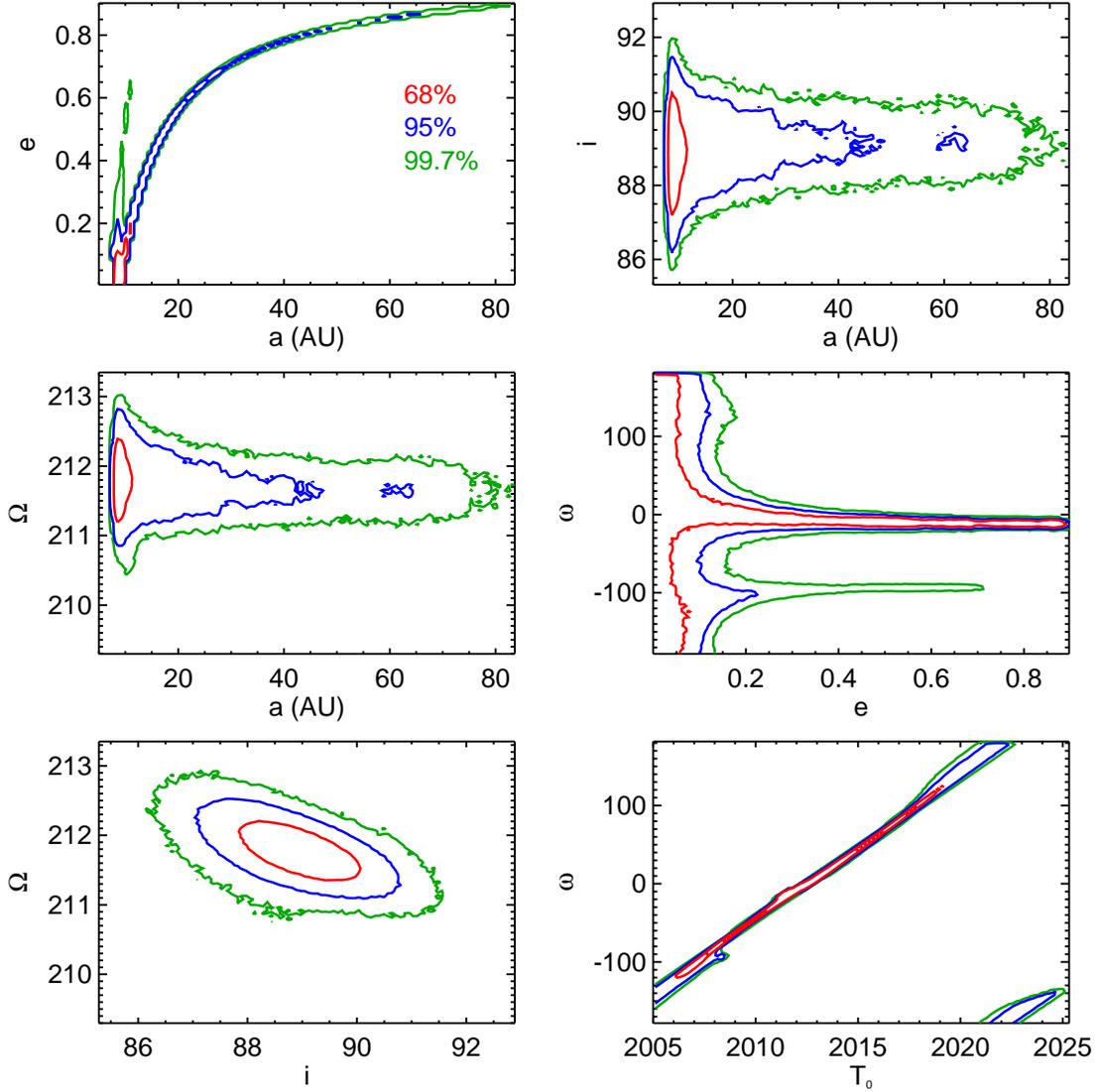}
\caption{Covariances between parameters (semi-major axis $a$, eccentricity 
$e$, inclination angle $i$, argument of periastron $\omega$, position angle 
of nodes $\Omega$, and epoch of periastron passage~$T_0$) for 
the MCMC orbital fit with total mass 
fixed to 1.75~M$_{\sun}$.  Lower semi-major axis corresponds to lower 
eccentricity, with the most probable orbits close to circular with small 
semi-major axis.  The position angle of nodes and inclination angle are 
tightly constrained and correlated, with position angle of nodes 
3.2$\sigma$ from the PA inner disk and 7.4$\sigma$ from the 
PA of the outer disk.
\label{contourfig}}
\end{figure}

Our results for the fixed-mass case 
show an  orbit with semi-major axis of about 9 AU 
is favored, with inclination angle and position angle of nodes (the two 
angles that describe the orientation of the orbit on the sky) very 
tightly constrained and eccentricity mainly $<$0.2.  Figure~\ref{contourfig} 
shows covariances between some orbital parameters 
for the fixed-mass fit.  There is a strong correlation between eccentricity 
and semi-major axis, with larger values of semi-major axis corresponding to 
more eccentric orbits.  Given the presence of the disk, it is unlikely that 
the orbit of the planet is significantly non-circular ($e \gtrsim$ 0.2), 
and we expect future 
observations to be most consistent with the shorter-period circular orbits 
from our MCMC chain.  With a uniform prior on eccentricity, we 
find $e<$ 0.15 at 68\% confidence and $e<$ 0.72 at 95\% confidence.  If we 
were to impose a prior that eccentricity must be smaller than 0.2, we 
would obtain $a$=8.9$^{+0.8}_{-0.3}$~AU and $P$=20.1$^{+2.5}_{-1.2}$~yr, 
with the distributions for the other parameters about the same as in 
the uniform prior case.  Our MCMC results find a median of time of 
maximum elongation (when the projected star-planet separation reaches a 
maximum) of 2012.63 and 68\% (95\%) confidence interval between 2012.55 
(2012.48) and 2012.70 (2012.79).  
Our Magellan epochs are after turn around time in 99\% of all 
orbits.  So while the Magellan data represent the largest separation between 
star and planet in our astrometric record, our orbit fitting results 
indicate that 
maximum elongation was reached prior to these data being taken.

Finally, we consider the floating-mass MCMC fit.  Though 
generally less constrained than the fixed-mass fit, the 
posterior mass distribution (lower right panel of Figure~\ref{histfig}) 
shows that the previous estimated mass of 1.75 M$_\sun$ is well within the 
68\% confidence interval.  While 
additional astrometry is required to place a more precise limit on the mass 
of the star $\beta$ Pic, this total mass distribution of 
1.6$\pm$0.3 M$_{\sun}$ 
indicates that our orbital fit and existing astrometric measurements are 
reasonable.  Semi-major axis and period are highly correlated in our 
floating-mass fit, and while the chains have not converged for 
semi-major axis and 
period individually (GR statistics of 1.2), the GR statistic for 
mass is 1.0012, indicating a reliable measurement of the mass.  If we were 
to impose the prior that $e<0.2$ in the floating-mass case, our mass 
constraints for $\beta$ Pic would slightly tighten to 1.7$\pm$0.2~M$_{\sun}$. 
Similarly, the semi-major axis range would shrink from 
24$^{+51}_{-15}$ AU to 9.0$^{+0.9}_{-0.4}$ AU.  The minimum $\chi^2$ reached 
in the floating-mass chain (41.64) is similar to that for the fixed-mass 
chain (41.67).

\citet{crifo_betapic} estimate the mass of $\beta$ Pic by 
comparing the position of the star on the HR diagram to theoretical 
evolutionary tracks, finding a good fit for masses between 1.7 and 1.8 
M$_{\sun}$.  \citet{blondel06} derive a similar mass, between 1.65 and 1.87 
M$_{\sun}$, again from evolutionary models and the HR diagram.  Our dynamical 
mass is independent of any evolutionary model and so provides independent 
confirmation of the mass of the star, and indicates these previous estimates 
were accurate.

\subsection{Comparison to Previous Fits}

We now compare our results to previously published MCMC orbital fits.  
Comparing our posteriors to the results from \citet{chauvin_orbit}, we have 
similar distributions for semi-major axis and eccentricity but tighter 
constraints on the viewing angles of the orbit.  For inclination angle 
and position 
angle of nodes, we find 88.9$\pm$0.7$^\circ$ and 211.8$\pm$0.3$^\circ$ 
compared to 88.5$\pm$1.7$^\circ$ and 212.6$\pm$1.5$^\circ$ from 
Chauvin et al.  We find similar distributions in semi-major axis 
and eccentricity, though with smoother posteriors indicating our 
MCMC chains are better converged.  (Chauvin et al. 
do not provide GR values for their fit but state that their 
GR statistics are consistent with convergence.)  Since 
the preferred orbits from our chains are close to circular (68\% having 
eccentricity less than 0.15), the argument of periastron and epoch of 
periastron 
passage are poorly defined as they were for Chauvin et al., since 
it is difficult to determine periastron in a near-circular orbit.  
Nevertheless, the two values are tightly correlated (bottom 
right panel of Figure~\ref{contourfig}).  While the location of 
periastron is not well-defined, the location of the planet at a particular 
epoch between $\sim$2000 and $\sim$2020 is.  

\citet{bonnefoy_orbit} refit the orbit 
with data from \citet{chauvin_orbit} as well as 
an additional VLT data point from 2012 and find similar results, with 
MCMC chains that appear more converged than those of \citet{chauvin_orbit}.  
The semi-major axis from this fit is reported to be within 8--10 AU at 
80\% confidence; we find a less constrained semi-major axis, with 80\% 
confidence lying between 8 and 14 AU.  We also find a wider 
range of eccentricities that fit the data, with 80\% confidence $e<0.35$ 
compared to $e<0.15$ reported by \citet{bonnefoy_orbit}.  Conversely, 
we find tighter constraints on the orientation of the orbit on the sky.  
From their Figure 8, we determine that their inclination angle distribution 
is centered on 88.7$^\circ$ with a FWHM of 3.5$^\circ$, and their 
distribution for position angle of nodes is centered on 212$^\circ$ and 
FWHM of 2.7$^\circ$.  Our distributions for these parameters have similar 
centers of 88.9$^\circ$ and 211.8$^\circ$, but with smaller FWHMs of 
1.7$^\circ$ and 0.7$^\circ$.  

After our paper was submitted, \citet{macintosh14} 
presented an additional measurement of the position of $\beta$ Pic b on 
18 November 2013 UT, measuring 0.434$\pm$0.006'' and 
211.8$\pm$0.5$^\circ$, using the Gemini Planet Imager (GPI) 
at Gemini-South.  
They fit this new datapoint along with the VLT astrometry presented in 
\citet{chauvin_orbit}.  Their astrometric coverage thus has a 3-year gap 
between early 2011 and late 2013, and a datapoint one year later than 
the orbital coverage we present here.  They find generally similar results for 
orientation of the orbit, $\Omega$ = 211.6$\pm$0.45$^\circ$ and $i$ = 
90.69$\pm$0.68$^\circ$ compared to 211.8$\pm$0.3$^\circ$ and 
88.9$\pm$0.7$^\circ$ we present here.  The peaks of their distributions for 
semi-major axis and eccentricity are similar to ours, but their 
posteriors do not reveal the same narrow degenerate 
tail toward large eccentricity and large semi-major axis.  (Note that 
\citealt{macintosh14} plot 10, 50, and 90\% confidence contours for their 
orbit fitting results, while 
we show 68.3, 95.4, and 99.7\% contours.)

When we fit their data (the combination of the 
\citealt{chauvin_orbit} and the GPI astrometry) with our MCMC method, we do 
not reproduce the posteriors of \citet{macintosh14} and still see the 
same long eccentricity tail.  In fact, the lowest chi-square orbit reached in 
the chain is in this tail ($\chi_\nu^2$ = 0.45), 
with semi-major axis 47.8 AU and 
eccentricity of 0.81.  The cause of this discrepancy is most likely the choice 
of priors, as a non-uniform prior on eccentricity can make high eccentricity 
orbits sufficiently unlikely so that the chain never leaves the low-period, 
low-eccentricity region of parameter space.  For other orbital 
parameters, including inclination angle (with a reflection about 90$^\circ$) 
and position angle of nodes, we precisely reproduce the 68\% confidence 
intervals given by \citet{macintosh14}.

Finally, we combine our dataset with the additional GPI 
astrometry point and re-fit the orbit.  We find similar 
parameters for the fixed-mass case as we present here, 
$a$ = 9.4$^{+11.8}_{-0.6}$ and $e$ = 
0.09$^{+0.49}_{-0.06}$, with the same degenerate tail toward long periods and 
large eccentricities as before.  We find slightly smaller error 
bars for inclination angle and position angle of nodes, 
$i$ = 89.1$\pm$0.6$^\circ$ and $\Omega$ = 
211.4$\pm$0.24$^\circ$.  While the 68\% confidence intervals are largely 
unchanged for these parameters, in the floating-mass case we significantly  
reduce the uncertainty on the mass of $\beta$ Pic, $M$ = 
1.76$^{+0.18}_{-0.17}$~M$_\sun$, reaching 10\% precision on the mass of the 
star.\footnote{Using our MCMC fit to the dataset 
presented here (VLT, NICI, and Magellan data) we would predict astrometry on 
18 November 2013 of 0.415$\pm$0.007'' and 212.3$\pm$0.6$^\circ$, within 1.5 
and 0.1 $\sigma$ of the GPI measurements, respectively.  For the floating 
mass case the predicted positions are 0.424$\pm$0.019'' and 
212.3$\pm$0.6$^\circ$.  The uncertainty 
in our predictions and the GPI measurement errors are similar for the 
fixed mass case, though with a slight offset in separation, so it is not 
surprising that adding the GPI datapoint to this analysis does not produce 
substantially different results.  For the floating mass case, however, the 
prediction uncertainty for separation is three times larger than the 
measurement error, so the new point does noticeably refine the fit.}

\section{Disk-Planet Alignment}

In order to compare the position angle of the orbit 
to the disk, we perform a custom reduction of our 2011 NICI 
data to recover the disk.  We begin by removing the azimuthally 
averaged profile of the point spread function 
from the individual images, as described 
in \citet{pipeline} except that the running azimuthal average used here is 
taken over 90 pixels instead of 30 pixels.  This process removes large 
scale ($>$90 pixels or 1.62'') azimuthal structure from both the stellar 
halo and the disk.  Since the $\beta$ Pic disk is known to be edge-on, this 
procedure does not alter the disk profile significantly.  After this 
step the disk reduction then proceeds with our standard ADI 
pipeline.

\begin{figure}
\epsscale{0.9}
\plotone{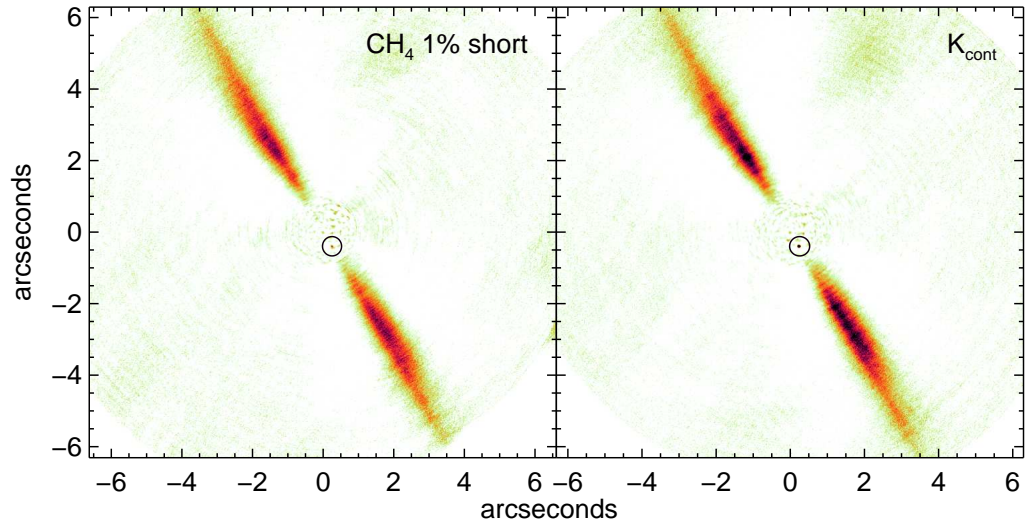}
\caption{Signal-to-noise maps of our Gemini NICI observations of $\beta$ Pic 
from 2011 
showing both disk and planet in $CH_4S$ (left) and $K_{cont}$ 
(right).  The figure is oriented with North up and East to the left.  The 
black circle marks the location of the planet.
\label{diskfig}}
\end{figure}

The signal-to-noise maps of the reduced images for the $CH_4L$ and 
$K_{cont}$ filters are shown in Figure~\ref{diskfig}.  The North/East (NE) 
and the South/West (SW) extensions of the disk were considered separately.  
As a function of radius we fit two Gaussians to the azimuthal disk profile, 
taking a 20 AU width at each sampling and stepping by 1 AU.  We compute the 
median position angles for the two sides of the outer disk 
(between 60 and 120 AU) of 
209.21$\pm$0.05$^\circ$ and 28.79$\pm$0.06$^\circ$.  The inner disk 
is only visible within 90 AU, and we find median values of the position 
angle between 60 and 90~AU of 212.99$\pm$0.07$^\circ$ and 
32.28$\pm$0.07$^\circ$.  These are errors in the fit only and do not 
include uncertainty in the position of north of 0.2$^\circ$, as we 
are concerned only with the relative offset between the orbital 
plane of the planet and the position angle of the disk.  We 
present these values along with relative and absolute errors in 
Table~\ref{table3}.

To ensure that 
our reduction process has not biased these measurements we simulate the disk 
as a series of rings given the fit to the data, and then create a simulated 
dataset with the same set of rotation angles in our ADI dataset (a total 
range of 42 degrees) to model self-subtraction.  The offset between 
the initial model and the measurements of the self-subtracted simulation 
was much smaller than 0.1$^\circ$, and so this effect is not the dominant 
source of error.  Rather it is residuals from the stellar point spread 
function that dominate the measurement of the disk position angle.

Our PA 
measurements are consistent with the cADI 2-component fit 
reported by \citet{lagrange12}, who find 29.07$^\circ$$^{+0.20}_{-0.19}$ 
and 209.00$^\circ$$^{+0.16}_{-0.15}$ for the outer disk.  Our measurements of 
offsets between outer and inner disk of 3.78$\pm$0.12$^\circ$ (SW) and 
3.49$\pm$0.13$^\circ$ (NE) are also close to the \citet{lagrange12} cADI value 
of 3.9$^{+0.6 \circ}_{-0.1}$.  Unlike \citet{lagrange12}, where the 
position of the center of the star is uncertain, the partially transmissive 
focal plane mask of NICI allows us to precisely measure the star position 
in images of the disk.

Using our position angle for the SW disk of 
209.21$\pm$0.05$^\circ$, our measurement of the position angle 
of nodes of the orbit of 211.8$\pm$0.3$^\circ$ is 7.4$\sigma$ discrepant with 
the position angle of the outer disk.  The inner warped disk has a 
position angle of 212.9$\pm$0.07$^\circ$, which is discrepant at the 
3.2$\sigma$ level.  Thus, we find the planet's orbital plane is not 
aligned with either disk.

\citet{currie_betapic} presented 
a preliminary orbit based on four astrometric points (the bare minimum needed 
for an orbit fit) and a non-MCMC method 
and found the position angle of nodes to be between the two disks.  When 
we apply our MCMC method to their data we find generally larger uncertainties 
in orbital parameters than \citet{currie_betapic} report.  Using an MCMC 
fit to their data, we find the position angle of nodes to be 
210.9$\pm$0.9$^\circ$, 1.8$\sigma$ greater than the outer disk PA and 
2.2$\sigma$ less than the inner disk PA.  \citet{chauvin_orbit} and 
\citet{bonnefoy_orbit} fit more data with a similar MCMC method to ours 
and find the orbital plane of the planet to be consistent 
with the inner disk ($\Omega$=212.6$\pm$1.5$^\circ$ and 
$\Omega$=212.0$\pm$1.1$^\circ$, respectively, both within 1$\sigma$ of 
the PA of the inner disk).  By including our higher precision astrometric 
data and longer time baseline, we find the position angle of nodes to be 
between the two disks, though closer to the inner disk.

\section{Discussion}

We now consider the implications of a misalignment between $\beta$ Pic b and 
the two disks.  Our observations recall a dynamical
picture of the $\beta$ Pic disk recently painted by \citet{dawson11}.  
If a planet on an inclined orbit (inclination $i_p$) is introduced into a 
disk of non-interacting particles with zero initial inclination, the secular 
theory of Laplace-Lagrange (e.g. \citealt{murray99}) show us that the 
inclinations of the particles will oscillate about $i_p$, creating 
a cuspy disk structure with apparent inclination $2i_p$ in its inner 
regions.  Our measurement that the planet's inclination is intermediate 
between the two observed disk planes seems to supports this picture.  
In contrast, earlier simulations by \citet{mouillet97} and 
\citet{augereau01} yielded a disk coplanar with the planet.

One important assumption built into the \citet{dawson11} model and 
supported by our observations is that the planetesimals are not yet 
collisionally relaxed 
(collisional relaxation in debris disks is the gradual 
process by which planetesimals lose energy and exchange momentum via 
collisions so that the distribution of their orbits approaches a steady 
state), as models of structure in other debris disks have 
assumed (e.g. \citealt{quillen06}, \citealt{rodigas14}).  
Another important assumption built into the \citet{dawson11} model 
is that the planet is introduced instantaneously, fully formed, at time 
zero.  We infer that the $\beta$ Pic disk is probably not 
collisionally relaxed; this inference should yield some interesting 
constraints on models of the collisional evolution of planetesimals in 
debris disks (e.g. \citealt{newvold13}).  Moreover, if the \citet{dawson11} 
model is indeed correct, it appears that the $\beta$~Pic planet 
was introduced to its current orbit suddenly compared to the secular time 
scale, perhaps scattered there by another planet.

This notion begs us to ponder the role of multiple planets in sculpting 
the disk. In the context of their model, \citet{dawson11} placed 
severe limits on the presence of a second planet in the system disturbing 
the disk.  In addition, \citet{absil13} exclude a second planet more massive 
than $\sim$5 M$_{Jup}$ outside of 0.2'' based on $L'$ high contrast images of 
$\beta$~Pic.  But what caused the inclination of $\beta$ Pic b if not an 
interaction with another massive planet?  And how did the planetesimals 
respond to this interaction?  These questions remain unanswered.

Finally, we consider the transit event noted by \citet{betapic_transit} 
in November 1981 and find very loose constraints on a transit of 
the planet given the available data.  We search for such a 
transit by finding the smallest projected distance between star and 
planet between 2015 and 2019 and then subtract one or two 
orbital period from the returned epoch for each step in our MCMC chains, 
choosing the epochs that are closest to 1981.  
We find the most recent closest approach to be at epoch 
2007.44$^{+0.14}_{-0.17}$ (crossing behind the star), with 
the next closest approach to take place in 2017.59$^{+0.60}_{-0.18}$, 
adopting the orientation of the orbit from the RV measurements of 
\citet{lagrange_rv} and \citet{snellen14} (see footnote 1).  
The median value for 
the epoch of closest approach one or two orbital period earlier than 2017 
is 1977.52, with a 68\% confidence interval between 1971.07 and 1982.14 
(95\% confidence between 1768 and 1993, given the long 
tail of the posterior for orbital period).  The 
smallest projected distance from star to planet is less than 
31 times the radius of $\beta$ Pic (1.8$\pm$0.2 R$_{\sun}$); 
\citealt{difolco04}) at 68\% confidence ($<$51 times the radius at 
95\% confidence).  Thus we cannot 
rule out a transit with the current orbital fit.  \citet{betapic_transit} 
and \citet{chauvin_orbit} speculate that the transit event may not be caused 
by the planet itself but rather by solid material entrained by the 
planet and carried in its Hill sphere.  We cannot rule out this possibility 
either, and more data are required.  We predict that the next transit  
will take place between 2017.41 and 2018.18 at 68\% confidence 
(2017.27--2019.19 at 95\% confidence), 
with a median epoch of 2017.59.  Additional astrometric observations in 
the next few years will provide a more precise transit window and 
transit probability.

\section{Conclusions}

We have examined the orbit of $\beta$ Pic b given five new epochs of 
data taken with Gemini/NICI and Magellan/MagAO, finding a 
semi-major axis of 9.1$^{+5}_{-0.5}$~AU and a period of 
21$^{+21}_{-2}$ years.  The astrometric record of $\beta$ Pic b is now long 
enough to be able to remove the assumption of the total system mass, which 
was needed by all previous fits to this orbit.  When we solve for the mass of 
$\beta$ Pic itself we find a value of 1.76$^{+0.18}_{-0.17}$~M$_{\sun}$, 
consistent with the expected value of 1.75~M$_{\sun}$.  The position angle 
of nodes for our fixed-mass orbit is offset from the observed position 
angles of the inner warped disk (at 3.2$\sigma$ significance) 
and from the outer disk (at 7.4$\sigma$ significance), 
suggesting that the disk is not collisionally relaxed.

Numerous degenerate orbital solutions exist for astrometric 
data that show minimal acceleration during the timeframe of the observations, 
in particular between semi-major axis, period, and eccentricity.  Observing 
significant acceleration, and in particular the reversal of direction at 
maximum elongation, greatly reduces these degeneracies.  
Our orbital fit indicates that the planet has 
reached maximum elongation and is currently moving back toward the star, 
crossing to the other side of the star by $\approx$2018.  
$\beta$~Pic~b has been 
observed extensively since its reappearance in 2009 
and the current window for studying the planet will remain open 
for just a few more years before the 
planet is undetectable behind the star again.  Advanced planet-finding 
instruments such as GPI and SPHERE will likely allow for orbital monitoring 
of the planet closer to the star, so the time of lost contact is likely to 
be significantly shorter than it was between 2003 and 2009.  The window 
for the next transit is between 2017.41 and 2018.18 at 68\% confidence, 
and future astrometric monitoring will provide a more precise prediction 
to guide photometric monitoring.

We thank Jessica Lu and Adam Kraus for helpful discussions.  B.A.B was in part 
supported by Hubble Fellowship grant HST-HF-01204.01-A awarded by
the Space Telescope Science Institute, which is operated by AURA for
NASA under contract NAS 5-26555.  This work was supported in part by NSF
grants AST-0713881 and AST-0709484 awarded to M. Liu.
The Gemini Observatory is operated by the Association of Universities for 
Research in Astronomy, Inc., under a cooperative agreement with the NSF on 
behalf of the Gemini partnership: the National Science Foundation (United 
States), the Science and Technology Facilities Council (United Kingdom), the 
National Research Council (Canada), CONICYT (Chile), the Australian Research 
Council (Australia), CNPq (Brazil), and CONICET (Argentina).  
This research has made use of the SIMBAD database,
operated at CDS, Strasbourg, France.  

{\it Facilities:} \facility{Gemini:South (NICI)},
\facility{Magellan II (MagAO+Clio2)},
\facility{Magellan II (MagAO+VisAO)}.

\bibliographystyle{apj}
\bibliography{apj-jour,betapic_orbit}
\clearpage

\begin{deluxetable}{lcccc}
\tabletypesize{\scriptsize}
\tablecaption{New Astrometry for $\beta$ Pic b\label{table1}}
\tablewidth{0pt}
\tablehead{
\colhead{Instrument} & \colhead{Filter} & \colhead{Sep. (``)} & 
\colhead{PA ($^\circ$)} & \colhead{Epoch (UT)} \\
}
\startdata
Gemini/NICI & $CH_4S$ (4\%) & 0.323$\pm$0.010 & 209.3$\pm$1.8 & 2009 Dec 03 \\
Gemini/NICI & $CH_4L$ (4\%) & 0.339$\pm$0.010 & 209.2$\pm$1.7 & 2009 Dec 03 \\
Gemini/NICI & $K_S$ & 0.407$\pm$0.005 & 212.9$\pm$1.4 & 2010 Dec 25 \\
Gemini/NICI & $CH_4S$ (1\%) & 0.455$\pm$0.003 & 211.9$\pm$0.4 & 2011 Oct 20 \\
Gemini/NICI & $K_{cont}$ & 0.452$\pm$0.005 & 211.6$\pm$0.6 & 2011 Oct 20 \\
Gemini/NICI & $CH_4S$ (1\%) & 0.447$\pm$0.003 & 210.8$\pm$0.4 & 2012 Mar 29 \\
Gemini/NICI & $K_{cont}$ & 0.448$\pm$0.005 & 211.8$\pm$0.6 & 2012 Mar 29 \\
Magellan/MagAO+Clio2 & 3.1 $\mu$m, 3.3 $\mu$m, $L'$, $M'$ & 0.461$\pm$0.014 & 211.9$\pm$1.2 & 2012 Dec 02 \\
Magellan/MagAO+VisAO & $Y_S$ & 0.470$\pm$0.010 & 212.0$\pm$1.2 & 2012 Dec 04 \\

\enddata
\end{deluxetable}

\begin{deluxetable}{lccccc}
\tabletypesize{\scriptsize}
\tablecaption{Orbital Parameters of $\beta$ Pic b\label{table2}}
\tablewidth{0pt}
\tablehead{
\colhead{Parameter} & \colhead{Lowest $\chi^2$} & \colhead{Median} & 
\colhead{68\% CI} & \colhead{95\% CI} & \colhead{GR Stat.} \\
}
\startdata
Fixed-Mass & & & & & \\
\tableline
Semi-major Axis (AU) & 9.6 & 9.1 & 8.6--14.4 & 8.2--48.0 & 1.0827 \\
Eccentricity                 & 0.08 & 0.08 & 0.02--0.40 & 0.00--0.82 & 1.0580 \\
Inclination Angle ($^\circ$) & 88.8 & 88.9 & 88.2--89.6 & 87.4--90.4 & 1.0006 \\
Argument of Periastron ($^\circ$) & 6.5 & -8 & -102--85 & -167--166 & 1.0022 \\
Position Angle of Nodes ($^\circ$) & 211.9 & 211.8 & 211.5--212.1 & 211.2--212.4 & 1.0011 \\
Epoch of Periastron Passage  & 2012.88 & 2012 & 2009--2019 & 2006--2023 & 1.0021 \\
Period (yr) & 22.60 & 21 & 19--41 & 18--251 & \ldots \\
\tableline
Floating-Mass & & & & & \\
\tableline
Semi-major Axis (AU) & 83.6 & 24.3 & 9.1--75.2 & 8.4--96.7 & 1.2532 \\
Eccentricity                 & 0.90 & 0.65 & 0.11--0.89 & 0.01--0.91 & 1.2473 \\
Inclination Angle ($^\circ$) & 89.2 & 89.0 & 88.3--89.7 & 87.6--90.3 & 1.0047 \\
Argument of Periastron ($^\circ$) & 347.4 & -16 & -46--1 & -133--116 & 1.0067 \\
Position Angle of Nodes ($^\circ$) & 211.6 & 211.7 & 211.4--211.9 & 211.2--212.3 & 1.0115 \\
Epoch of Periastron Passage  & 2011.52 & 2011 & 2010--2013 & 2006--2021 & 1.0012 \\
Period (yr) & 581.53 & 96 & 22--516 & 18--758 & 1.2282 \\
Mass (M$_{\sun}$) & 1.73 & 1.64 & 1.37--1.95 & 1.13--2.38 & 1.0012\\
\enddata
\tablecomments{Results from the MCMC fit to the orbit of $\beta$
Pic b and orbital parameters for the lowest $\chi^2$ within the 68\% confidence
interval from the chains (Figure~\ref{orbitfig}).  The final column gives 
the Gelman-Rubin statistic for each parameter: values closer to unity 
indicate a higher degree of convergence.}
\end{deluxetable}

\begin{deluxetable}{lccc}
\tabletypesize{\scriptsize}
\tablecaption{Disk Measurements of $\beta$ Pic\label{table3}}
\tablewidth{0pt}
\tablehead{
\colhead{Disk Component} & \colhead{PA ($^\circ$)} & \colhead{Relative Error ($^\circ$)} & \colhead{Absolute Error ($^\circ$)} \\
}
\startdata
NE Outer (60--120 AU) & 28.79 & 0.06 & 0.3 \\
SW Outer (60--120 AU) & 209.21 & 0.05 & 0.3 \\
NE Inner (60--90 AU) & 32.28 & 0.07 & 0.3 \\
SW Inner (60--90 AU) & 212.28 & 0.07 & 0.3 \\
\enddata
\end{deluxetable}

\end{document}